\documentclass{moriond}

\def\be{\begin{equation}}
\def\ee{\end{equation}}
\def\bea{\begin{eqnarray}}
\def\eea{\end{eqnarray}}

\newcommand{\Photo}{}

\begin{document}
\vspace*{4cm}
\title{AXION MASS IN THE CASE OF POST-INFLATIONARY PECCEI-QUINN SYMMETRY BREAKING}

\author{ A. RINGWALD}

\address{Deutsches Elektronen-Synchritron DESY, Notkestr. 85, D-22607 Hamburg, Germany}

\maketitle\abstracts{
The axion not only solves the strong $CP$ puzzle, but it also may be the main constituent of cold dark matter.
We review the axion dark matter predictions for the case that the Peccei-Quinn symmetry  is restored after inflation.
}

\section{\boldmath Axion solution of the strong $CP$ puzzle and the axion mass}

Already in the early days of Quantum Chromodynamics (QCD) it was realised that the most generic 
Lagrangian of QCD contained also a term of the form
${\mathcal L}_{\rm QCD} \supset -  
\frac{\alpha_s}{8\pi}\,  \bar\theta \,G_{\mu\nu}^b \tilde{G}^{b,\mu\nu}$,
where $\alpha_s$ is the strong coupling, $G_{\mu\nu}^b$ is the gluonic field strength, $\tilde{G}^{b,\mu\nu}$
its dual, and $\bar\theta \in [-\pi,\pi]$ an angular parameter. This term violates parity ($P$) and time-reversal ($T$) 
invariances 
and, due to the $CPT$ conservation theorem,  also $CP$ invariance. Consequently, it induces $CP$ violation in 
flavor-diagonal strong interactions,
notably non-zero electric dipole moments of nuclei. However, none have been detected up to date. The best constraint
currently comes from the electric dipole moment of the neutron, which is bounded by $|d_n|<  2.9\times 10^{-26} e$\,cm. 
A comparison with the
prediction, $d_n \sim e \bar\theta m^\ast_q/m_n \sim 6\times 10^{-17}e$\,cm, where $m^\ast_q \equiv m_u m_d/(m_u+m_d)$ is the reduced quark and $m_n$ the neutron mass,  leads to the conclusion that $|\bar\theta |< 10^{-9}$. 
This is the strong $CP$ puzzle. 

In Peccei-Quinn (PQ) extensions~\cite{Peccei:1977hh} of the Standard Model (SM), the symmetries of the latter 
are extended by a global $U(1)_{\rm PQ}$ symmetry which is 
spontanously broken by the vacuum expectation value (VEV) 
 of a new complex singlet scalar 
field,  $\langle{|\sigma |^2}\rangle=v_{\rm PQ}^2/2$, which is assumed to be much larger than the Higgs VEV. 
SM quarks or new exotic quarks are supposed to carry PQ charges such that
$U(1)_{\rm PQ}$ is also broken by the gluonic triangle anomaly, 
$\partial_\mu J_{U(1)_{\rm PQ}}^\mu \supset 
-\frac{\alpha_s}{8\pi}\,N_{\rm DW}\, G_{\mu\nu}^a \tilde G^{a\,\mu\nu}$, 
where $N_{\rm DW}$ is a model-dependent integer. 
Under these circumstances and at energies above the confinement scale $\Lambda_{\rm QCD}$ 
of QCD, but far below $v_{\rm PQ}$, the PQSM
reduces to the SM plus a pseudo Nambu-Goldstone boson~\cite{Weinberg:1977ma,Wilczek:1977pj} -- the axion $A$ -- 
whose field, $\theta (x) \equiv A(x)/f_A\in [-\pi,\pi]$, corresponding to the angular degree of freedom
of $\sigma$, acts as a space-time dependent $\bar\theta$ 
parameter, 
${\mathcal L}_\theta \supset 
\frac{f_A^2}{2} \,\partial_\mu \theta \partial^\mu \theta
- \frac{\alpha_s}{8\pi}\,\theta(x)\,G_{\mu\nu}^c {\tilde G}^{c,\mu\nu}$, 
with $f_A \equiv v_{\rm PQ}/N_{\rm DW}$.
Therefore, the $\overline\theta$-angle  can be eliminated by a shift $\theta (x) \to \theta (x) -\overline\theta$.  
At energies below $\Lambda_{\rm QCD}$, 
the effective potential of the shifted field, which for convenience we again denote by $\theta(x)$, will then coincide 
with the vacuum energy of QCD as a function of $\overline\theta$, which, on general 
grounds, has an absolute
minimum at $\theta =0$, implying that there is no strong $CP$ violation: $\langle \theta\rangle =0$.  In particular, 
$V(\theta ) = \frac{1}{2} \chi \theta^2 + {\mathcal O}(\theta^4) $,
where $\chi\equiv \int d^4x\, \langle q(x)\,q(0)\rangle$, with  $q(x)\equiv \frac{\alpha_s}{8\pi}\,G_{\mu\nu}^c(x) {\tilde G}^{c,\mu\nu}(x)$, is the topological susceptibility. 
A recent lattice determination found~\cite{Borsanyi:2016ksw} 
$\chi = [75.6(1.8)(0.9) {\rm MeV}]^4$, which agrees well with the result from NLO chiral perturbation theory~\cite{diCortona:2015ldu},  
$\chi = [75.5(5) {\rm MeV}]^4$, leading to the following prediction of the axion mass in terms of the 
axion decay constant $f_A$,
\begin{equation}
\label{zeroTma}
m_A\equiv \frac{1}{f_A}\sqrt{\frac{d^2 V}{d\theta^2}}{|_{\theta = 0}}= \frac{\sqrt{\chi}}{f_A} = 
57.0(7)\,   \left(\frac{10^{11}\rm GeV}{f_A}\right)\mu \textrm{eV}. 
\end{equation}

\section{Axion cold dark matter in the case of post-inflationary PQ symmetry breaking}

In a certain range of its decay constant, the axion not only solves the strong $CP$ puzzle, but is also a cold dark matter
candidate~\cite{Preskill:1982cy,Abbott:1982af,Dine:1982ah}. The extension of this range depends critically on the cosmological history. It is particularly constrained in the case on which we concentrate here: post-inflationary PQ symmetry restoration and subsequent breaking.\footnote{Remarkably, this case is strongly favored in the case of 
saxion (modulus of $\sigma$) or saxion/Higgs inflation~\cite{Ballesteros:2016xej}.} 

In the early universe, after the PQ phase transition, the axion field takes on 
random initial values in domains of the size of the causal horizon. 
Within each domain, the axion field evolves according to  
\begin{equation}
\label{KG}
\ddot \theta + 3 H(T) \dot\theta  + \frac{\chi(T)}{f_A^2} \sin\theta= 0 ,
\end{equation}
with temperature dependent Hubble expansion rate $H(T)\sim T^2/M_P$ and topological 
susceptibility~\cite{Pisarski:1980md} $\chi (T)\propto T^{-(7 + 3/n_f)}$,  for temperatures far above the QCD 
quark-hadron crossover, $T_c^{\rm QCD}\simeq 150$\,MeV ($n_f$ is the number of active quark flavors).  
At very high temperatures, $v_{\rm PQ} > T\gg T_c^{\rm QCD}$, the Hubble friction term  is much larger than the potential term in (\ref{KG}), $3 H(T)\gg \sqrt{\chi(T)}/f_A$,  and the axion field is frozen at its initial value. At temperatures around a GeV, 
however, when $\sqrt{\chi(T)}/f_A \simeq 3 H(T)$, the field starts to 
evolve towards the minimum of the potential and to oscillate around the $CP$ conserving ground state. 
Such a spatially coherent oscillation has an equation of state 
like cold dark matter, $w_A \equiv p_A/\rho_A \simeq 0$ (here $p_A$ and $\rho_A$ are the pressure and the 
energy density of the axion field, respectively). 
Averaging over the initial values of the axion field in the many domains filling our universe -- at temperatures around 
a GeV the size of a domain is around a mpc  --  one 
obtains~\cite{Borsanyi:2016ksw,Ballesteros:2016xej}  
 $\Omega_A^{\rm (VR)}h^2 = 
(3.8\pm  0.6   )\times 10^{-3} \,\left(f_A \over { 10^{10}\, {\rm GeV}}\right)^{1.165}$,
for the fractional contribution of axion cold dark matter to the energy density of the universe 
from this so-called vacuum realignment (VR) mechanism~\cite{Preskill:1982cy,Abbott:1982af,Dine:1982ah}.  
Here, the exponent, $1.165$, arises from the temperature dependence of  
$\chi(T)$ at $T\sim $\,GeV, which has recently been determined quite precisely 
from lattice QCD~\cite{Borsanyi:2016ksw}. Requiring, that the axion dark matter abundance 
should not exceed the observed one, this result implies a lower limit on the axion 
mass~\cite{Borsanyi:2016ksw}: 
\begin{equation}
m_A > 28(2)\,\mu{\rm eV} \,.
\end{equation}

However, so far we have neglected that the domain-like structure discussed above comes along with a network of one and two dimensional topological defects -- strings~\cite{Davis:1986xc} and domain walls~\cite{Sikivie:1982qv} -- which are formed at the boundaries of the domains. Their collapse will also produce axions. 

Axion strings are formed at the same time when the domain-like structure appears, cf. at the PQ phase transition. 
 In the string cores, of typical radius $1/m_\rho$, where $m_\rho \equiv \sqrt{2\lambda_\sigma} v_{\rm PQ}$ is the mass of the saxion (the particle excitation of the saxion field),  
topology hinders the breaking of the PQ symmetry and a huge energy density 
is stored. As the network evolves, the overall string length decreases by straightening and collapsing loops. 
Moreover, some energy is radiated in the form of low-momentum axions.  The energy density in the network of global strings is expected to reach a scaling behaviour,  
$\rho_{\rm S} = \zeta \frac{\mu_{\rm S}}{t^2}$,
with string tension $\mu_{\rm S} \equiv \pi v_{\rm PQ}^2 \ln\left(\frac{m_\rho t }{\sqrt{\zeta}}\right)$,
where $\zeta$ is independent of time. 
This scaling behavior implies that the number density of axions radiated from strings (S) can be 
estimated as 
\begin{equation}
\label{Nastring}
{n_{A}^{\rm (S)}(t)} 
\simeq
\frac{\zeta}{\epsilon}\frac{v_{\rm PQ}^2}{t}\left[\ln\left(\frac{m_\rho t}{\sqrt{\zeta}}\right)-3\right] ,
\end{equation}
where the dimensional parameter $\epsilon$ gives a measure of the average energy of the radiated axions in units
of the Hubble scale, $\epsilon \equiv \langle E_A\rangle/(2\pi /t)$. 
A number of field theory simulations have indicated that the network of strings evolves indeed toward the scaling solution
with $\zeta = {\mathcal O}(1)$ and $\epsilon ={\mathcal O}(1)$. 
The latter value implies that most of the axions produced from strings become non-relativistic during the radiation-dominated era and contribute to the cold dark matter abundance. Adopting the values~\cite{Kawasaki:2014sqa} 
$\zeta = 1.0 \pm 0.5$ and $\epsilon = 4.02 \pm 0.70$, one finds from (\ref{Nastring}) for the contribution of strings to today's dark matter abundance~\cite{Ballesteros:2016xej} 
$\Omega_A^{\rm (S)} h^2
\approx 
7.8^{+6.3}_{-4.5} \times 10^{-3} \times N_{\rm DW}^2 
\left( \frac{f_A}{10^{10}\ {\rm GeV}}\right)^{1.165}$,
where the upper and lower end correspond to the maximum and minimum values obtained by using the above 
error bars on $\zeta$ and $\epsilon$. They do not take into account a possible large theoretical error due to the fact that 
the field theory simulations can only be performed at values of $\ln ( m_\rho t )\sim$\,a few, much smaller than the realistic value, $\sim 50$, and thus require an extrapolation.  

Domain walls appear at temperatures of the order of a GeV, when the axion field, in any of the causally connected domains at this epoch, relaxes into one of the $N_{\rm DW}$ distinct but degenerate minima of the effective potential 
effective potential, $V(A,T) = \chi (T) \left[ 1- \cos ( N_{\rm DW} A/v_{\rm PQ} )\right]$, in the 
 interval $-\pi v_{\rm PQ}\leq A \leq +\pi v_{\rm PQ}$. Between the domains, there appear two dimensional 
topological defects dubbed domain walls whose thickness and stored energy density is controlled by $\chi (T)$.  Importantly, strings are always attached by $N_{\rm DW}$ domain walls,
due to the fact that the value of the phase of the PQ field $\sigma$ must vary from $-\pi$ to $\pi$ around the string core. 
Therefore, hybrid networks of strings and domain walls, so-called string-wall systems, are formed at $T={\mathcal O}(1)$\,GeV. 
Their evolution strongly depends on the model-dependent value of $N_{\rm DW}$. 

For $N_{\rm DW} = 1$, strings are pulled by one domain wall, which causes the disintegration into smaller pieces of a wall bounded by a string~\cite{Vilenkin:1982ks}.
String-wall systems are short-lived in this case, and their collapse (C) contributes an amount~\cite{Kawasaki:2014sqa} 
$\Omega_A^{\rm (C)} h^2
\approx 
3.9^{+2.3}_{-2.1} \times 10^{-3} \times
\left( \frac{f_A}{10^{10}\ {\rm GeV}}\right)^{1.165}$  
to dark matter, 
resulting in a total abundance
\begin{equation}
\Omega_A h^2 
\approx \left( \Omega_A^{\rm{(VR)}} + \Omega_A^{\rm{(S)}} + \Omega_A^{\rm{(C)}}\right) h^2
\approx 1.6^{+1.0}_{-0.7}\times 10^{-2}\times \left(\frac{f_A}{10^{10}\,\mathrm{GeV}}\right)^{1.165}.
\label{omega_a_tot_short}
\end{equation}
Therefore, in post-inflationary PQ symmetry breaking models with $N_{\rm DW}=1$, the axion may explain all of cold dark matter in the universe
if its decay constant and mass are in the range 
\begin{equation} 
\label{mass_range}
f_A \approx (3.8-9.9)\times 10^{10}\,{\rm GeV}\hspace{3ex}  \Leftrightarrow\hspace{3ex} 
m_A \approx (58 - 150)\ \mu{\rm eV}\,.
\end{equation}

This prediction, however, has recently been challenged by the results from a new field theory simulation 
technique 
designed to work directly at high string tension with  $\ln (m_\rho t)\sim 50$ and 
to treat vacuum realignment, string, and string-wall contributions in a unified way~\cite{Klaer:2017ond}. 
The reported dark matter axion mass, 
\begin{equation}
m_A=(26.2 \pm 3.4)\,\mu{\rm eV}\,,
\end{equation} 
where the error now only includes the uncertainty from $\chi(T)$, 
is significantly lower than (\ref{mass_range}). It indicates that axions from strings and walls are negligible,   
despite of the fact that the string networks appear to have a higher energy density ($\zeta \sim 4$) than those observed in conventional field theoretic simulations ($\zeta\sim 1$). This implies that the produced axions have a larger energy,  $\epsilon \sim 40$, 
and that dynamics at smaller scales -- outside the range of applicability of the new simulation method~\cite{Klaer:2017ond} -- can be relevant for the determination of the axion DM abundance.
Further studies on the dynamics of string-wall systems are required to include precise modelling of physics at smaller distance scales.

Fortunately, there are new axion dark matter direct detection experiments aiming to probe 
the mass region of interest for $N_{\rm DW}=1$ models with post-inflationary PQ symmetry breaking, notably CULTASK~\cite{Chung:2016ysi}, HAYSTAC~\cite{Zhong:2018rsr}, and 
MADMAX~\cite{TheMADMAXWorkingGroup:2016hpc}.  

For $N_{\rm DW} > 1$, the string-wall systems are stable, since the strings are pulled in 
$N_{\rm DW}$ different directions. The existence of such stable domain walls is firmly excluded by standard cosmology~\cite{Zeldovich:1974uw}. Stability can be avoided if there exist further interactions which explicitly break the 
PQ symmetry, e.g. 
${\mathcal L} \supset 
g M_P^4 \left(\frac{\sigma}{M_P} \right)^N 
 +h.c.$,  
where $g$ is a complex dimensionless coupling, $M_{P}$ 
is the reduced Planck mass, and $N$ is an integer ($>4$). The appearance of such terms is motivated by the fact that global symmetries are not protected from effects of quantum gravity. 
They give rise to an additional contribution in the low energy effective potential of the axion field, which lifts the degeneracy of the minima of the QCD induced potential by an amount~\cite{Ringwald:2015dsf} 
$\Delta V \simeq -2 |g| M_P^4 \left(\frac{v_{\rm PQ}}{\sqrt{2}M_P} \right)^N \left[
\cos \left( \frac{2\pi N}{N_{\rm DW}} + \Delta_D \right) - \cos \Delta_D \right]
$,
where $\Delta_D = \arg(g) - N \overline{\theta}$, 
and acts like a volume pressure on domain walls. 
If $\Delta V$ is small, domain walls live for a long time and emit a lot of axions, potentially overclosing the universe. 
On the other hand, if $\Delta V$ is large, it shifts the location of the minimum of the axion effective potential and leads to large $CP$ violation, spoiling the axionic solution of the strong $CP$ problem. 
A detailed investigation of the parameter space exploiting the results of field theory simulations~\cite{Kawasaki:2014sqa} 
showed~\cite{Ringwald:2015dsf} that there exists a valid region in parameter space if 
$N = 9$ or $10$.\footnote{The absence of PQ symmetry breaking operators with $4<N<9$ can be naturally explained if the PQ symmetry arises accidentally as a low energy remnant 
from a more fundamental 
discrete symmetry~\cite{Ringwald:2015dsf,Ernst:2018bib}.} In the case of $N_{\rm DW}=6$ and $N=9\,(10)$, and allowing a mild tuning of $|g|$,  
the axion can explain the observed dark matter abundance
for
\begin{equation} 
4.4\times 10^7\,(1.3\times 10^9)\,{\rm GeV} < f_A < 1\times 10^{10}\,{\rm GeV}\  \Leftrightarrow  \ 
0.56\,{\rm meV} <   m_A <  130\,(4.5)\,{\rm meV}\, .
\end{equation}

Intriguingly, a DFSZ axion ($N_{\rm DW}=6$) in such a mass range can explain the accumulating hints of excessive energy losses of stars in various stages of their evolution~\cite{Giannotti:2017hny}. 
In this range, axion dark matter direct detection may be difficult, but not impossible~\cite{Horns:2012jf,Baryakhtar:2018doz}.
Fortunately, it is aimed to be probed by the fifth force experiment ARIADNE~\cite{Arvanitaki:2014dfa} and the helioscope 
IAXO~\cite{Armengaud:2014gea}.

\end{document}